\documentclass[a4paper,11pt]{article}

\usepackage{amsmath}
\usepackage{siunitx}
\usepackage{graphicx}

\usepackage{hyperref}
\usepackage{amsmath,amsthm,amssymb}
\usepackage{xcolor}
\usepackage{epstopdf}
\usepackage{overpic}
\usepackage{bbold}
\usepackage{textcomp}
\usepackage{physics}

\usepackage{authblk}

\providecommand{\keywords}[1]
{
  \small
  \textbf{\textit{Keywords---}} #1
}

\renewcommand{\vec}[1]{\boldsymbol{#1}}
\newcommand{\dtau}{\,\text{d}\tau}
\newcommand{\ds}{\,\text{d}\vec{s}}
\newcommand{\D}[1]{\,\text{d} #1}

\begin{document}

\title{Hybrid FFT algorithm for fast demagnetization field calculations on non-equidistant magnetic layers}

\author[1]{Paul Heistracher \thanks{paul.thomas.heistracher@univie.ac.at}}
\author[1]{Florian Bruckner}
\author[1]{Claas Abert}
\author[1]{Christoph Vogler}
\author[1]{Dieter Suess}

\affil[1]{Christian Doppler Laboratory of Advanced Magnetic Sensing and Materials, Faculty of Physics, University of Vienna, Austria}

\maketitle

\begin{abstract}
In micromagnetic simulations, the demagnetization field is by far the computationally most expensive field component and often a limiting factor in large multilayer systems.
We present an exact method to calculate the demagnetization field of magnetic layers with arbitrary thicknesses. 
In this approach we combine the widely used fast-Fourier-transform based circular convolution method with an explicit convolution using a generalized form of the Newell formulas.
We implement the method both for central processors and graphics processors and find that significant speedups for irregular multilayer geometries can be achieved.
Using this method we optimize the geometry of a magnetic random-access memory cell by varying a single specific layer thickness and simulate a hysteresis curve to determine the resulting switching field.
\end{abstract}

\newpage
\section{Introduction}
Micromagnetic simulations are becoming an increasingly important design tool to support the development of magnetic devices
such as magnetic random access memory (MRAM) devices \cite{duine_synthetic_2018, makarov_cmos-compatible_2016} or magnetic sensing devices \cite{suess_topologically_2018}.
Due to miniaturization, these devices enter the domain of micromagnetics where sub-micrometer magnetic structures such as domain walls or vortices become relevant.
These structures are yet large enough such that the atomic structure can be neglected in a continuum approximation and the relevant physics can be properly described in the micromagnetic model.

In the micromagnetic model the demagnetization field is the only global interaction in the system and by far the computationally most expensive one.
An efficient calculation of the demagnetization field is a topic of intense research
and there are multiple approaches to be found in literature.
Well established methods calculate the demagnetization field using a FFT-based fast convolution with a point-wise tensor-vector multiplication in Fourier space \cite{hayashi_calculation_1996, fabian_three-dimensional_1996}. Scalar potential methods also use FFT-based convolutions but reduce to a point-wise tensor-scalar product in Fourier space \cite{berkov_solving_1993,abert_fast_2012}.
Other methods include Fourier-transform methods on irregular grids \cite{kritsikis_fast_2008}, fast-multipole methods \cite{blue_using_1991,seberino_concise_2001}, non-uniform grid methods \cite{livshitz_nonuniform_2009} and tensor-grid methods \cite{juselius_parallel_2007, exl_fast_2012, exl_non-uniform_2014}.
From all these methods the FFT-based convolution method with tensor-vector multiplication in Fourier-space is arguably the most widely used as it is the default method 
in the established finite-difference micromagnetic codes of the OOMMF project \cite{m._j._donahue_oommf_1999}, mumax3 \cite{vansteenkiste_design_2014} and fidimag \cite{bisotti_fidimag_2018}.
This approach, however, restricts the discretization to an equidistant mesh which is often unpractical when simulating multilayer structures with experimentally given non-equidistant layer thicknesses.

To address this issue we present a hybrid FFT algorithm calculating the demagnetization field of magnetic layers with arbitrary thicknesses.
A similar method using mesh transfer was recently published \cite{lepadatu_efficient_2019}.
Our algorithm is hybrid in the sense that we still perform an equidistant FFT-based fast convolution along the
two axes of the layers and an explicit convolution along the third axis allowing non-equidistant thicknesses.
For this method we analytically derive the demagnetization tensor for cuboids of arbitrary shape
by extending the Newell formulas \cite{newell_generalization_1993}.
We implement the hybrid FFT algorithm for graphical processing units (GPUs) and central processing units (CPU) and investigate the scaling of the method as function of the number of layers.

We find that this method can be more efficient in cases where the non-equidistant layer thicknesses can not properly be discretized by an equidistant mesh and where the number of layers is not excessively high. This is the case for many experimental multilayer devices such as synthetic antiferromagnets and giant magnetoresistance sensors.
The ability to vary respective layer thicknesses also gives rise to the application of optimization routines uncommon in micromagnetic finite-difference codes.
With our GPU-accelerated implementation we use Newton iteration to optimize the geometry of an MRAM cell by varying a single layer thickness of the stack in order to obtain a minimum average demagnetization field in the free layer.

\section{Generalized Newell Equation}
In current-free regions the magnetic field fulfills $\vec{\nabla} \times \vec{h} = \vec{0}$. Thus it can be expressed as the gradient field $\vec{h} = -\vec{\nabla} \phi$, where the scalar potential $\phi$ is determined by
\begin{equation}
	\phi(\vec{r}) = \frac{1}{4 \pi} \int_{\tau'} \vec{M}(\vec{r}') \cdot \vec{\nabla}' \left( \frac{1}{\abs{\vec{r} - \vec{r}'}} \right) \dtau'
	.
\end{equation}
Assuming a homogeneous magnetization $\vec{M}$ inside of the source region $\tau'$ the average magnetic field $\langle \vec{h} \rangle_\tau$ inside of the target region $\tau$ can be expressed as
\begin{equation}
	\langle \vec{h} \rangle_\tau = -\frac{1}{\tau} \int_\tau \vec{\nabla} \phi(\vec{r}) \dtau = -\vec{M} \cdot \vec{N}
	,
\end{equation}
with the demagnetization tensor $\vec{N}$ defined by
\begin{align} \label{eqn:demag_tensor}
\begin{split}
\vec{N} &= \frac{1}{4 \pi \tau} \int_\tau \dtau \, \vec{\nabla} \int_{\tau'} \dtau' \, \boldsymbol{\nabla}' \frac{1}{\abs{\vec{r} - \vec{r}'}} \\
       &= \frac{1}{4 \pi \tau} \oint_{\partial \tau} \ds \oint_{\partial \tau'} \ds' \frac{1}{\abs{\vec{r} - \vec{r}'}}
       .
\end{split}
\end{align}
In the following Eqn.~\eqref{eqn:demag_tensor} should be calculated for a rectangular source region $\tau'$ with dimensions $(\Delta x', \Delta y', \Delta z')$ and a rectangular target region $\tau$ with dimensions $(\Delta x, \Delta y, \Delta z)$. The offset of the target region is given by $(X,Y,Z)$. Each component of the demagnetization tensor requires to calculate the interaction between two pairs of rectangular surfaces. We denote the components of the demagnetization tensor as

\begin{equation}
\vec{N}= \left(
\begin{matrix}
N_{xx} & N_{xy} & N_{xz}\\
N_{yx} & N_{yy} & N_{yz}\\
N_{zx} & N_{zy} & N_{zz}
\end{matrix}\right)
\end{equation}

and compute the components $N_{xx}$ and $N_{xy}$ in the following. The remaining components can then be obtained by permutation of the variables.

\subsection{Component $N_{xx}$}
\begin{figure}
	\centering
	\includegraphics{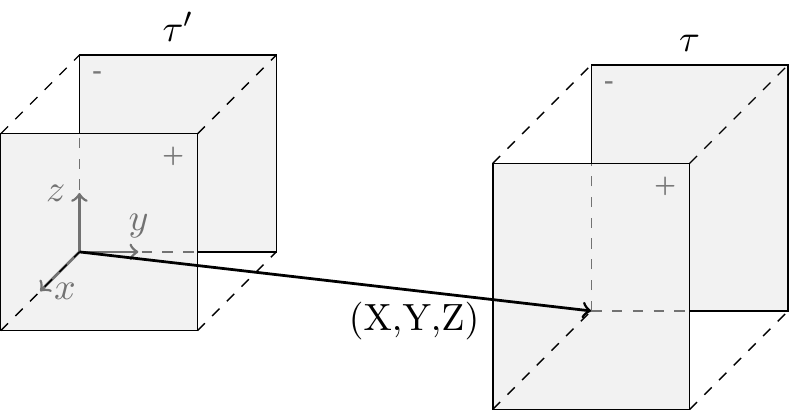}
	\caption{Rectangular surfaces which contribute to the $N_{xx}$ component of the demagnetization tensor. The signs indicate whether the outward normal points in positive or negative $x$-direction.}
	\label{fig:Nxx}
	%The positive sign indicates an outward normal facing in positive $x$-direction and a negative sign indicates an outward normal facing in negative $x$-direction, respectively.}
\end{figure}

The $N_{xx}$ component can be obtained by a sum of integrals involving only surfaces with normal vector in $x$-direction (see Fig.~\ref{fig:Nxx}):
\begin{align}\label{eqn:Nxx}
\begin{split}
N_{xx}(X,Y,Z) = \frac{1}{4 \pi \abs{\tau}} [ &F(X, Y, Z) \\
                                           - &F(X + \Delta x, Y, Z) \\
                                           - &F(X - \Delta x', Y, Z) \\
                                           + &F(X + \Delta x - \Delta x', Y, Z) ],
\end{split}
\end{align}
where $F(X,Y,Z)$ describes the interaction between two parallel faces with offset $(X,Y,Z)$ and $\abs{\tau} = \Delta x \, \Delta y \, \Delta z$ and reads
\begin{equation}\label{eqn:F}
	F(X, Y, Z) = \int\limits_Z^{Z+\Delta z} \D{z} \int\limits_Y^{Y+\Delta y} \D{y} \int\limits_0^{\Delta z'} \D{z'} \int\limits_0^{\Delta y'} \D{y'} \frac{1}{\sqrt{X^2+(y-y')^2+(z-z')^2}}
	.
\end{equation}
Substituting $\tilde{y} = y-y'$ and $\tilde{z} = z-z'$ into Eqn.~\eqref{eqn:F} and adapting the integration limits accordingly leads to the simpler expression
\begin{align}\label{eqn:F_simplified}
F(X, Y, Z) = \int\limits_Z^{Z+\Delta z} \D{z} \int\limits_Y^{Y+\Delta y} \D{y} \int\limits_{z-\Delta z'}^z \D{\tilde{z}} \int\limits_{y-\Delta y'}^y \D{\tilde{y}} \; \frac{1}{\sqrt{X^2+\tilde{y}^2+\tilde{z}^2}},
\end{align}
which in turn can be split into 16 integrals $F_2$ of the form (see Appendix \ref{sec:F2_splitting})
\begin{align}\label{eqn:F2}
F_2(X, Y, Z) = \int\limits_0^Z \D{z} \int\limits_0^Y \D{y} \underbrace{\int\limits_0^z \D{\tilde{z}} \int\limits_0^y \D{\tilde{y}} \; \frac{1}{\sqrt{X^2+\tilde{y}^2+\tilde{z}^2}}}_{\tilde{f}(y,z)}.
\end{align}
Note that $F_2(X,Y,Z)$ is independent of the size of source and target region and thus it can be adopted from \cite{newell_generalization_1993} without modification, where it is defined as
\begin{align}\label{eqn:F_2}
F_2(X, Y, Z) = f(X,Y,Z) - f(X,0,Z) - f(X,Y,0) + f(X,0,0),
\end{align}
where $f$ is the indefinite integral of $\tilde{f}$
\begin{align}
\begin{split}
f(x,y,z) &= \frac{y}{2} \; (z^2-x^2) \; \sinh^{-1}\left(\frac{y}{\sqrt{x^2+z^2}}\right) \\
         &+ \frac{z}{2} \; (y^2-x^2) \; \sinh^{-1}\left(\frac{z}{\sqrt{x^2+y^2}}\right) \\
         &- xyz \; \tan^{-1} \left(\frac{yz}{xR} \right) + \frac{1}{6} \; (2x^2-y^2-z^2) \; R
\end{split}
\end{align}
and $R=\sqrt{x^2+y^2+z^2}$.
Due to the symmetry considerations the last three terms in equation (\ref{eqn:F_2}) cancel out and as a result the component $N_{xx}$ can be expressed in 64 instead of 256 terms. 

\subsection{Component $N_{xy}$}

\begin{figure}
	\centering
	\includegraphics{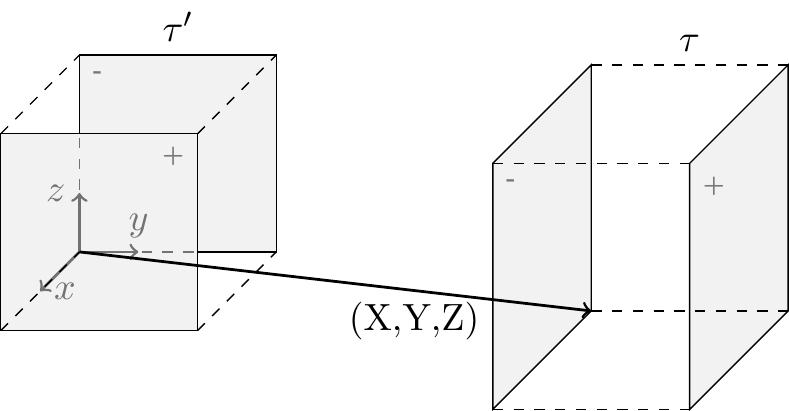}
	\caption{Rectangular surfaces which contribute to the $N_{xy}$ component of the demagnetization tensor.}
	\label{fig:Nxy}
\end{figure}

The $N_{xy}$ component can be split into four interactions between source-planes with normal vector in $y$-direction and target-planes with normal vector in $x$-direction (see Fig.~\ref{fig:Nxy}):
\begin{align}\label{eqn:Nxy}
\begin{split}
N_{xy}(X,Y,Z) = \frac{1}{4 \pi \abs{\tau}} [ &G(X, Y, Z) \\
											- &G(X - \Delta x', Y, Z) \\
											- &G(X, Y + \Delta y, Z) \\
											+ &G(X - \Delta x', Y - \Delta y, Z) ],
\end{split}
\end{align}
where $G(X,Y,Z)$ describes the interaction between two orthogonal faces with offset $(X,Y,Z)$ and $\abs{\tau} = \Delta x \, \Delta y \, \Delta z$ and reads
\begin{align}\label{eqn:G}
G(X, Y, Z) = \int\limits_Z^{Z+\Delta z} \D{z} \int\limits_X^{X+\Delta x} \D{x} \int\limits_0^{\Delta z'} \D{z'} \int\limits_0^{\Delta y'} \D{y'} \frac{1}{\sqrt{x^2+(Y-y')^2+(z-z')^2}}.
\end{align}
Similar to the previous case we can rewrite equation (\ref{eqn:G}) by substituting $\tilde{y}=Y-y'$ and $\tilde{z}=z-z'$ and adapting the integration limits.
This yields

\begin{align}\label{eqn:G_transfromed}
G(X, Y, Z) = \int\limits_Z^{Z+\Delta z} \D{z} \int\limits_X^{X+\Delta x} \D{x} \int\limits_{z-\Delta z'}^{z} \D{\tilde{z}} \int\limits_{Y-\Delta y'}^{Y} \D{\tilde{y}} \frac{1}{\sqrt{x^2+\tilde{y}^2+\tilde{z}^2}},
\end{align}
which also can be spit up into 16 integrals $G_2$ of the form (see Appendix \ref{sec:G2_splitting})

\begin{equation}\label{eqn:G2}
	G_2(X,Y,Z)=\int\limits_{0}^{Z}\D{z} \underbrace{\int\limits_{0}^{X}\D{x} \int\limits_{0}^{z}\D{\tilde{z}} \int\limits_{0}^{Y}\D{\tilde{y}} \frac{1}{\sqrt{x^2+\tilde{y}^2+\tilde{z}^2}}}_{\tilde{g}(z)}.
\end{equation}
Note that $G_2(X,Y,Z)$ is again independent of the size of the source and target region and can be adopted from \cite{newell_generalization_1993} and reads
\begin{equation}\label{eqn:G2_as_g}
	G_2(X,Y,Z)=g(X,Y,Z) - g(X,Y,0)
	,
\end{equation}
where $g$ is the indefinite integral of $\tilde{g}$:
\begin{align}
\begin{split}
	g(x,y,z) = & \ 
	 xyz \ \textmd{sinh}^{-1} \Bigg(\frac{z}{\sqrt{x^2 + y^2}}\Bigg)\\
	&+ \frac{y}{6} (3z^2-y^2) \ \textmd{sinh}^{-1} \Bigg(\frac{x}{\sqrt{y^2 + z^2}}\Bigg)\\
	&+ \frac{x}{6} (3z^2-x^2) \ \textmd{sinh}^{-1} \Bigg(\frac{y}{\sqrt{x^2 + z^2}}\Bigg)\\
	&-\frac{z^3}{6} \ \textmd{tan}^{-1} \Big(\frac{xy}{zR}\Big) - \frac{zy^2}{2} \ \textmd{tan}^{-1} \Big(\frac{xz}{yR}\Big)\\
	&-\frac{zx^2}{2} \ \textmd{tan}^{-1} \Big(\frac{yz}{xR}\Big) - \frac{xyR}{3}
	.
\end{split}
\end{align}
Due to the symmetry considerations the second term in equation (\ref{eqn:G2_as_g}) cancels out. As a result, the component $N_{xy}$ can be expressed in 64 instead of 128 terms.

\section{Demagnetization field in non-equidistant finite-differences}
The demagnetization field of a given magnetic material can be calculated by the convolution of the demagnetization tensor $\vec{N}$ with the normalized magnetization field $\vec{m}$ and reads

\begin{equation}
    \vec{H}(\vec{r}) = -M_s \int_\Omega \vec{N}(\vec{r}-\vec{r}')\vec{m}(\vec{r}') d\vec{r}',
\end{equation}
where $M_s$ is the saturation magnetization of the magnetic material and $\vec{m} = \vec{M}/M_s$ is the normalized magnetic field with $|\vec{m}| = 1$.
When considering a discrete distribution of the magnetization field $\{\vec{m_i}\}$ at points $\{\vec{i}\}$ the field at $\vec{i}$ can be expressed as
\begin{equation}\label{eqn:Hi_discrete}
\vec{H_i} = -M_s \sum_{\vec{j}} \vec{N}_{\vec{i} - \vec{j}} \cdot \vec{m_j},
\end{equation}
where $\vec{i}$ and $\vec{j}$ are multi-indices with $\vec{i} = (i_1, i_2, i_3)$ and $\vec{j} = (j_1, j_2, j_3)$ and the sum goes over all points including $\vec{i}$.

The demagnetization tensor connecting $\vec{i}$ with $\vec{j}$ can be written as
\begin{equation}
	N_{\mathbf{i}-\mathbf{j}} = \frac{1}{V_i} \iint_{\Omega_\text{ref}} N\left( \sum_k (i_k-j_k) \Delta \mathbf{r}_k + \mathbf{r} - \mathbf{r}'\right) d\mathbf{r} d\mathbf{r}'
	.
\end{equation}

With the adapted Newell equations presented above we can compute the field generated by two homogeneously magnetized rectangular cuboids where each cuboid is allowed to have an arbitrary size.
In the following we consider the common finite-difference model in which we divide the magnetic material into an array of rectangular cuboids. We restrict the cuboids to have common dimensions of $\Delta x$ and $\Delta y$ in the $x$- and $y$-directions, respectively. As of the $z$-direction, we discuss two separate cases. The first case we assume an equal dimension $\Delta z$ for all cuboids.
As all layers have the same thickness we refer to this case as the \textit{equidistant} case. In the second case, we assume a constant $z$-dimension for each layer in the $xy$-plane but allow each individual layer $i$ to have an arbitrary thickness $\Delta z_i$. This is referred to as the \textit{non-equidistant} case.
In order to calculate the micromagnetic demagnetization field we apply FFTs along the $x$-, $y$- and $z$- directions for the \textit{equidistant}. In the \textit{non-equidistant} case we perform FFTs only in the $x$- and $y$- directions and
explicitly perform the convolution along the $z$-direction.

\subsection{Discrete convolution theorem}
With the discrete convolution theorem one can express the convolution as a cell-wise multiplication in Fourier space
\begin{equation}
	\mathcal{F}(f*g) = \mathcal{F}(f)\mathcal{F}(g),
\end{equation}
where $\mathcal{F}$ denotes the discrete Fourier transform and $f*g$ is the discrete convolution given by 
\begin{equation}
	(f*g)_i = \sum_j f_j \ g_{i-j}
	.
\end{equation} 
Using scalar indices and exchanging the convolution operation with the tensor-vector multiplication one can write equation (\ref{eqn:Hi_discrete}) as
\begin{equation}
	 H_{k,i_1,i_2,i_3} = - M_s \sum_l \sum\limits_{j_1,j_2,j_3} N_{kl,i_1-j_1,i_2-j_2,i_3-j_3} \, m_{l,j_1,j_2,j_3},
\end{equation}
where the free index $k$ runs over the three components of the vector field.
Applying the discrete convolution theorem for the fist two dimensions this becomes

\begin{align}
	H_{k,i_1,i_2,i_3} 
	&= - M_s \sum\limits_l \mathcal{F}_1^{-1} \sum\limits_{j_2,j_3} \left[ \mathcal{F}_1(N_{kl})_{i_2-j_2,i_3-j_3} \, \mathcal{F}_1(m_l)_{j_2,j_3} \right]_{i_1}\\
	&= - M_s\sum\limits_l \mathcal{F}_2^{-1} \mathcal{F}_1^{-1} \sum\limits_{j_3} \left[ \mathcal{F}_1 \mathcal{F}_2(N_{kl})_{i_3-j_3} \, \mathcal{F}_1 \mathcal{F}_2(m_l)_{j_3} \right]_{i_1,i_2}\label{eq:FFT_v2}.
\end{align}
For the non-equidistant case the remaining convolution in the third dimension has to be calculated explicitly
as the discrete convolution theorem is not applicable along this axis.
In the equidistant case, in contrast, equation (\ref{eq:FFT_v2}) further reduces to

\begin{equation}\label{eq:FFT_3D}
	H_{k,i_1,i_2,i_3} = - M_s\sum\limits_l \boldsymbol{\mathcal{F}}^{-1} \left[ \boldsymbol{\mathcal{F}} (N_{kl}) \, \boldsymbol{\mathcal{F}}(m_l) \right]_{i_1,i_2,i_3},
\end{equation}
where $\boldsymbol{\mathcal{F}}=\mathcal{F}^1\mathcal{F}^2\mathcal{F}^3$ is the three dimensional Fourier transform.

\subsection{Scaling of the method}
\begin{figure}[h!]
    \centering
    \includegraphics{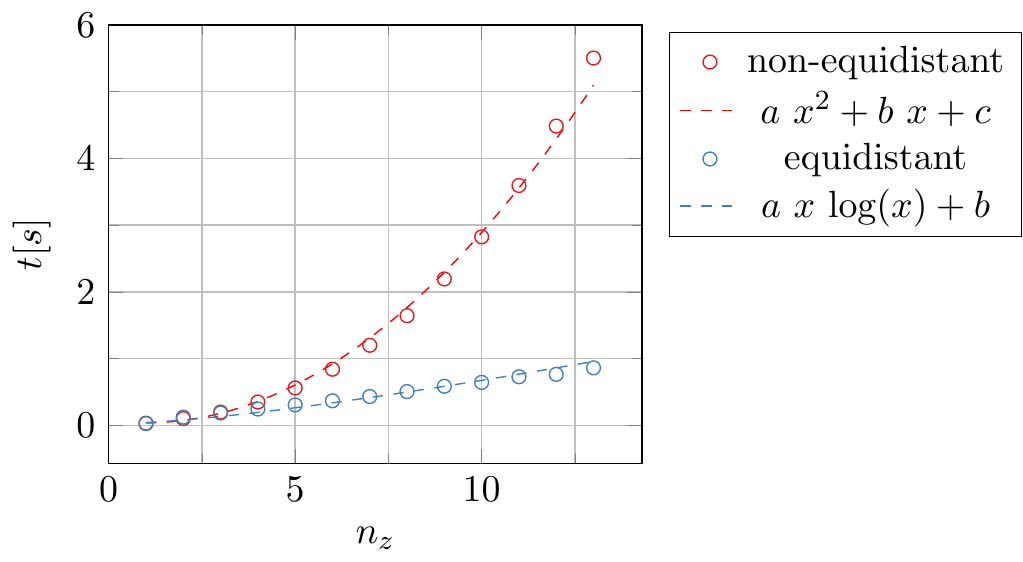}
    \caption{Evaluation time of the demagnetization field as function of $n_z$. Circles denote the measured average CPU time and dashed lines represent the fit-functions given in the legend.}
    \label{fig:timing_cpu}
\end{figure}

The fast convolution method asymptotically scales with $\mathcal{O}(n \ \textmd{log} \ n)$ in each dimension where $n$ is the number of nodes in the respective dimension. In contrast, an explicit convolution scales with $\mathcal{O}(n^2)$.
Accordingly, when we consider a fixed number of cells $n_x$ and $n_y$ in $x$- and $y$-direction and $n_z$ cells in $z$-direction, the presented \textit{non-equidistant} method scales with $\mathcal{O}(n_z^2)$, whereas the \textit{equidistant} method scales with $\mathcal{O}(n_z \ \textrm{log} \ n_z)$.
Figure \ref{fig:timing_cpu} shows the computation time of the demagnetization field as a function of $n_z$ for constant $n_x = n_y = 256$.
The measured timings are depicted as circles and represent the average value of 1000 field evaluations. The dashed lines represent nonlinear least squares fits and are obtained using the functions $f(x) = a \ x^2 + b \ x + c$ for the \textit{non-equidistant} case and $f(x) = a \ x \ \textrm{log}(x) + b$ for the \textit{equidistant} case. The Marquardt-Levenberg-algorithm is used for fitting and includes the standard deviation of the measurement points. Compared to the expected scaling the fits show good agreement.

\begin{figure}[h!]
    \centering
    \includegraphics{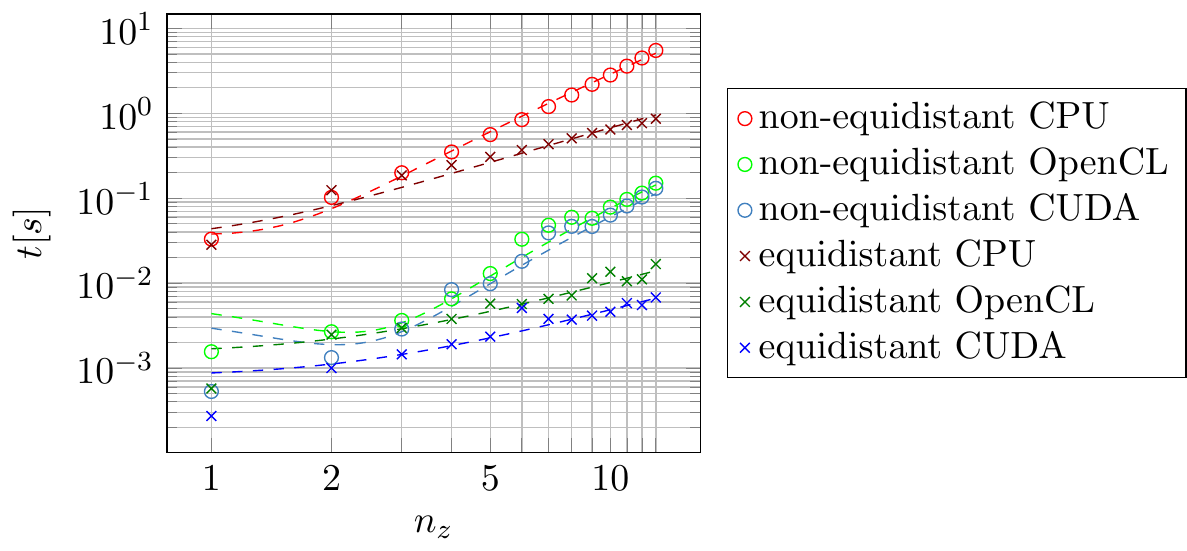}
    \caption{Evaluation time of the demagnetization field as function of $n_z$ comparing the non-equidistant and equidistant method for CPU, OpenCL and CUDA implementations. Dashed lines represent the two respective fit-functions as given in Figure \ref{fig:timing_cpu}.}
    \label{fig:timing_full}
\end{figure}

For the implementation of the two methods the general purpose GPU library ArrayFire is used which allows the usage of CPU, CUDA$^{\textmd{\textregistered}}$ and OpenCL\texttrademark \ backends \cite{yalamanchili_arrayfire_2015}.
The timings shown in Figure \ref{fig:timing_cpu} are single core CPU measurements performed on a AMD Ryzen\texttrademark \ 7 1700X processor.
In Figure \ref{fig:timing_full} we compare these CPU numbers both with CUDA and OpenCL timings performed on a NVIDIA$^{\textmd{\textregistered}}$ Tesla $^{\textmd{\textregistered}}$ V100 PCIe 16GB graphics card and observe substantial speedup due to the usage of GPU hardware. The dashed lines again represent data fits and the same respective fit functions as described above are used for the \textit{non-equidistant} and for the \textit{equidistant} case.
The parameters $a, b$ and $c$ differ for each fit function.

The \textit{non-equidistant} method can lead to significant simulation speedups for systems where the layers have different thicknesses which can not be properly discretized by an equidistant mesh
as is often the case in the simulation of magnetic multilayer systems.
Moreover, the \textit{non-equidistant} discretization naturally allows the variation of specific layer thicknesses which opens up new possibilities for geometry optimization.
In the next section we present the application of such an optimization procedure for the design of an magnetic random  access memory (MRAM) cell.

\newpage
\section{Layer thickness optimization}
\begin{figure}
    \centering
    \includegraphics{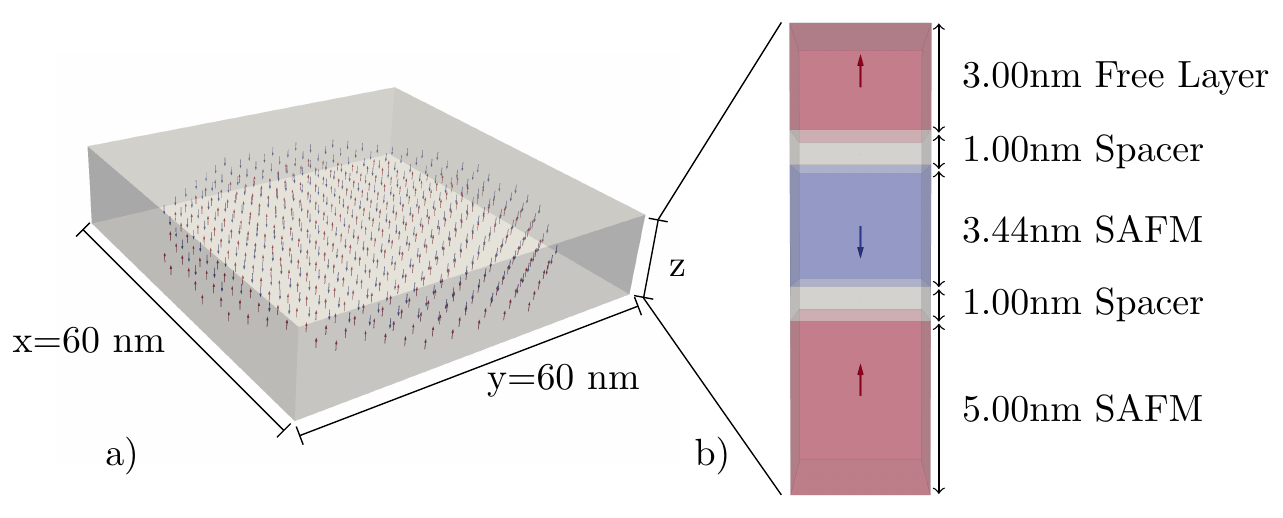}
    \caption{a) Cylindrical synthetic antiferromagnet geometry with diameter of \SI{60}{\nano\meter}. b) Dimensions of the different layers. The thickness of the middle layer is obtained by the Newton method minimizing the average $z$-component of the demagnetization field in the free layer.
    }
    \label{fig:stack}
\end{figure}

The use of the \textit{non-equidistant} method is not only a convenient way to simulate given irregular layer thicknesses, but also allows the variation of specific layer dimensions in optimization problems.
Especially in multilayer systems many properties such as the stray field magnitude can be tuned by varying individual layer thicknesses.
In the following,
we consider a synthetic antiferromagnet (SAFM) consisting of two layers and are interested in minimizing the average $z$-component of the demagnetization field these layers generate in a third magnetic layer which we refer to as the free layer.
We consider cylindrical layers as shown in Figure \ref{fig:stack} a).
The diameter of the system is \SI{60}{\nano\meter} and we use a discretization of $n_x = n_y = 64$ along the $x$- and $y$-axis as well as $n_z=5$ layers in the $z$-direction.

The first layer has a thickness of \SI{5}{\nano\meter} and a pinned magnetization field pointing in positive $z$-direction. 
The layer above is a \SI{1}{\nano\meter} thick non-magnetic spacer layer.
The middle layer is a magnetic layer with a pinned magnetization in negative $z$-direction and an initial thickness of \SI{5}{\nano\meter}. 
This value is then varied by applying the Newton method and becomes \SI{3.44}{\nano\meter}.
This is followed by another non-magnetic spacer layer with an thickness of \SI{1}{\nano\meter}.
The last layer is the magnetic free layer with a thickness of \SI{3}{\nano\meter}.

\begin{figure}
    \centering
    %\raggedleft
    \includegraphics{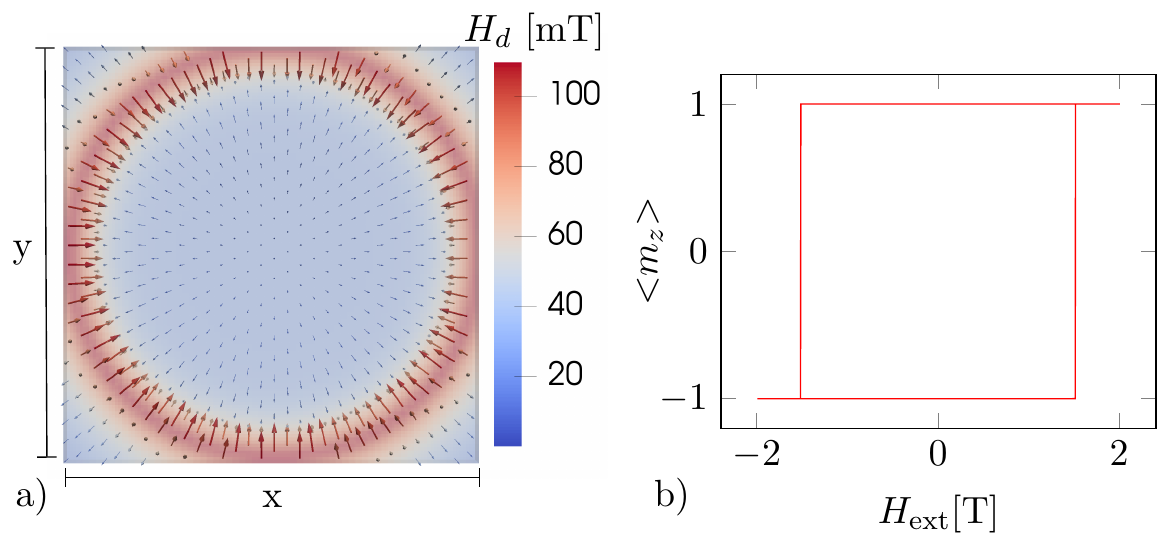}
	\caption{a) Demagnetization field in the free layer with minimal average $m_z$-component as obtained by optimizing the SAFM layer thickness. The field strength is color encoded, arrows indicate field orientation and magnitude. b) Easy axis hysteresis loop showing the MRAM cell switching process as obtained by micromagnetic energy minimization.}
	\label{fig:h_demag_freelayer}
\end{figure}

The optimized layer geometry obtained by Newton iteration is indicated in Figure \ref{fig:stack} b).
The newton iteration is terminated after seven steps as the value of
average $z$-component of the demagnetization field
approached zero up to the fifth decimal place which is around micromagnetic precision.
The obtained demagnetization field in the free layer is shown in \ref{fig:h_demag_freelayer} a). The circular shape is a result of the cylindrical stack layout.

With the optimized SAFM layer thickness we perform a full micromagnetic simulation to investigate the switching process of the free layer. Therefore we pin the magnetization of the two SAFM layers and apply an external field in the free layer.
For the micromagnetic parameters we use values for CoFe as in \cite{suess_topologically_2018}.
Accordingly, we assume a saturation magnetization of $J_s = \mu_0 M_s = \SI{1.75}{T}$, an exchange constant of $A_\textmd{ex}=\SI{1.5e-11}{J/\meter}$ and additionally an uni-axial anisotropy of $K_\textmd{u}=\SI{2.09e6}{J/\meter^3}$ in positive $z$-direction.
For the hysteresis loop we apply an external field along the $z$-direction and use an limited-memory Broyden–Fletcher–Goldfarb–Shanno energy minimization algorithm to relax the magnetization field for each applied external field magnitude.
Figure \ref{fig:h_demag_freelayer} b) shows the obtained hysteresis loop when using $2000$ discrete field-steps yielding a switching field of $\SI{1.51}{T}$.

\section{Conclusion}
We propose an explicit method for efficiently calculating the demagnetization field in magnetic multilayer systems.
By extending the Newell equations and performing an explicit convolution of non-equidistant layers
this method is a convenient way to discretize irregular layer thicknesses and can lead to a significant reduction in computation time compared to the equidistant method.
The formulation is well suited to be parallelized for GPU hardware, allowing additional speedups.
We demonstrate that both our CPU and GPU implementations yield the expected scaling
and highlight the possibility of varying single layer thicknesses with the non-equidistant method using simple optimization methods such as Newton iteration.
This can be of great practical use and is used to optimize a MRAM cell stack geometry by minimizing the average strayfield magnitude in the free layer.

\section{Acknowledgments}
The financial support by the Austrian Federal Ministry for Digital and Economic Affairs and the National Foundation for Research, Technology and Development is gratefully acknowledged.

\section{Competing Interests}
The authors declare no competing financial interests.

\appendix
\section{Transformation of $F(X,Y,Z)$} \label{sec:F2_splitting}
The 4-fold integral \eqref{eqn:F_simplified} can be split into 16 part-integrals with the lower integration limits equal to zero. At first one only considers the integrals over the $y$ and $\tilde{y}$ variables and transforms the inner integral limits accordingly:
\begin{align}
\begin{split}
\int\limits_Y^{Y+\Delta y}\!\!\!\!\!\!\D{y} \int\limits_{y-\Delta y'}^y\!\!\!\!\!\!\D{\tilde{y}} \ h(\tilde{y}) &= \int\limits_Y^{Y+\Delta y}\!\!\!\!\!\!\D{y} \int\limits_0^y\!\!\!\D{\tilde{y}} \ h(\tilde{y}) - \int\limits_Y^{Y+\Delta y}\!\!\!\!\!\!\D{y} \int\limits_0^{y-\Delta y'}\!\!\!\!\!\!\D{\tilde{y}} \ h(\tilde{y}) \\
                                                                    &= \int\limits_Y^{Y+\Delta y}\!\!\!\!\!\!\D{y} \int\limits_0^y\!\!\!\D{\tilde{y}} \ h(\tilde{y}) - \int\limits_{Y-\Delta y'}^{Y+\Delta y-\Delta y'}\!\!\!\!\!\!\!\!\!\!\D{\bar{y}} \ \int\limits_0^{\bar{y}}\!\!\!\D{\tilde{y}} \ h(\tilde{y})
                                                                    ,
\end{split}
\end{align}
where the function $h(\tilde{y})$ is used as a placeholder for the integrand. In the last step we substituted $y$ with $\bar{y}$ to shift the inner integration limit.
The outer integral can directly be split into two parts, which results in 4 normalized terms.
Applying the same procedure to the $z$ and $\tilde{z}$ variables finally yields the 16 normalized $F_2$ terms.
For sake of a better readability one can group the 16 terms into four sets of four by introducing a new function $F_1$ (note that here one groups positive and negative $y$ offsets, whereas in \cite{newell_generalization_1993} only the positive offsets of $y$ and $z$ where grouped):
\begin{align}
\begin{split}
F_1(X,Y,Z) &= F_2(X, Y+\Delta y,           Z) \\
           &- F_2(X, Y,                    Z) \\
           &- F_2(X, Y+\Delta y-\Delta y', Z) \\
           &+ F_2(X, Y-\Delta y',          Z).
\end{split}
\end{align}
Putting everything together yields:
\begin{align}
\begin{split}
F(X,Y,Z) &= F_1(X, Y, Z+\Delta z          ) \\
         &- F_1(X, Y, Z                   ) \\
         &- F_1(X, Y, Z+\Delta z-\Delta z') \\
         &+ F_1(X, Y, Z-\Delta z'         ).
\end{split}
\end{align}

\section{Transformation of $G(X,Y,Z)$} \label{sec:G2_splitting}
In a similar manner we can express $G(X,Y,Z)$ in equation (\ref{eqn:G_transfromed} as 16 part-integrals with lower integration limits of zero.
We start from equation (\ref{eqn:G2}) and split the integral for the $z$ and $\tilde{z}$ variables and transform the inner integration limit:
\begin{align}
\begin{split}
\int\limits_{Z}^{Z+\Delta z}\!\!\!\!\!\!\D{z}
\int\limits_{z -\Delta z'}^{z} \!\!\!\!\!\!\D{\tilde{z}} \ h(\tilde{z})
	&= 
	\int\limits_{Z}^{Z+\Delta z}\!\!\!\!\!\!\D{z}
	\int\limits_0^{z} \!\!\D{\tilde{z}} \ h(\tilde{z}) 
	-
	\int\limits_{Z}^{Z+\Delta z} \!\!\!\!\!\!\D{z}
	\int\limits_0^{z - \Delta z'} \!\!\!\D{\tilde{z}} \ h(\tilde{z}) \\
	&=
	\int\limits_{Z}^{Z+\Delta z}\!\!\!\!\!\!\D{z}
	\int\limits_0^{z} \!\!\!\D{\tilde{z}} \ h(\tilde{z})
	-
	\int\limits_{Z- \Delta z'}^{Z + \Delta z - \Delta z'}\!\!\!\!\!\!\!\!\!\!\!\D{\bar{z}} \ \
	\int\limits_0^{\bar{z}} \!\!\D{\tilde{z}} \ h(\tilde{z}).
\end{split}
\end{align}
The function $h(\tilde{y})$ again is used as a placeholder for the integrand.
The outer integrals can be split into two parts straightforwardly. For the $x$ and $\tilde{y}$ variables we apply the same procedure and obtain 16 integrals.
Introducing $G_1(X,Y,Z)$ we sort them into four by four terms:

\begin{equation}
\begin{split}
G_1(X,Y,Z) &=G_2(X, Y, Z + \Delta z) \\
&- G_2(X, Y, Z)  \\
&- G_2(X, Y , Z + \Delta z - \Delta z') \\
&+ G_2(X, Y , Z - \Delta z').
\end{split}
\end{equation}

Finally, we can write

\begin{equation}
\begin{split}
G(X,Y,Z) 
	&= G_1(X + \Delta x, Y, Z) \\
	&- G_1(X + \Delta x, Y - \Delta y', Z) \\
	&- G_1(X, Y, Z) \\
	&+ G_1(X, Y- \Delta y', Z).
\end{split}
\end{equation}

\bibliographystyle{ieeetr}
\bibliography{refs}

\end{document}